\documentclass[10pt,conference]{IEEEtran}
\IEEEoverridecommandlockouts
\usepackage{cite}
\usepackage{amsmath,amssymb,amsfonts}
\usepackage{bbding}
\usepackage{algorithmic}
\usepackage{graphicx}
\usepackage{textcomp}
\usepackage{xcolor}
\usepackage{verbatim}
\usepackage{pifont}
\usepackage{fancyvrb}
\usepackage{fancyhdr}
\usepackage{colortbl}
\usepackage{hhline}
\usepackage{tabularx}
\usepackage{subfig}
\usepackage{url} 
\usepackage{times}
\usepackage{latexsym}
\usepackage[T1]{fontenc}
\usepackage[utf8]{inputenc}
\usepackage{microtype}
\usepackage{inconsolata}
\usepackage{graphicx}
\usepackage{listings}
\usepackage{booktabs}
\definecolor{lightgray}{gray}{0.9}
\lstdefinelanguage{pattern}{
  language=Python,
    backgroundcolor=\color{gray!10!white},   
    numberstyle=\tiny\color{gray},
    basicstyle=\ttfamily\footnotesize,
    breakatwhitespace=false,   
    keywordstyle=\color{black},
    breaklines=false,   
    breakatwhitespace=false,  
    numbersep=5pt,                  
    showspaces=false,                
    showstringspaces=false,
    showtabs=false
}

\begin{document}

\title{Can GPT-O1 Kill All Bugs? \\ {An Evaluation of GPT-Family LLMs on QuixBugs}}

\author{
Haichuan Hu\textsuperscript{\rm 1}, 
  Ye Shang\textsuperscript{\rm 2}, 
  Guolin Xu\textsuperscript{\rm 3}, 
  Congqing He\textsuperscript{\rm 4}, 
  {\bf Quanjun Zhang}\textsuperscript{\rm 2 \thanks{\ Corresponding author.}}
  \\
  \textsuperscript{\rm 1}Alibaba Cloud \\
  \textsuperscript{\rm 2}State Key Laboratory for Novel Software Technology, Nanjing University, China\\
  \textsuperscript{\rm 3}School of Computer Science and Technology, Chongqing University, China \\
  \textsuperscript{\rm 4}University Sains Malaysia \\
  huhaichuan.hhc@alibaba-inc.com; 
  \{quanjun.zhang,201250032\}@smail.nju.edu.cn; \\
  xuguolin0626@gmail.com;
  hecongqing@hotmail.com;
}
\maketitle

\begin{abstract}
LLMs have long demonstrated remarkable effectiveness in automatic program repair (APR), with OpenAI's ChatGPT being one of the most widely used models in this domain.
Through continuous iterations and upgrades of  GPT-family models, their performance in fixing bugs has already reached state-of-the-art levels.
However, there are few works comparing the effectiveness and variations of different versions of GPT-family models on APR.
In this work, inspired by the recent public release of the GPT-o1 models, we conduct the first study to compare the effectiveness of different versions of the GPT-family models in APR. 
We evaluate the performance of the latest version of the GPT-family models (i.e., O1-preview and O1-mini), GPT-4o, and the historical version of ChatGPT on APR.
We conduct an empirical study of the four GPT-family models against other LLMs and APR techniques on the QuixBugs benchmark from multiple evaluation perspectives, including repair success rate, repair cost, response length, and behavior patterns.
The results demonstrate that O1's repair capability exceeds that of prior GPT-family models, successfully fixing all 40 bugs in the benchmark. 
Our work can serve as a foundation for further in-depth exploration of the applications of GPT-family models in APR.
\end{abstract}

\section{Introduction}
Automated program repair (APR) is widely regarded as one of the most important and challenging areas in software quality assurance~\cite{zhang2023survey}.
In the literature, numerous APR techniques have been proposed to fix software bugs automatically.
Among them, traditional template-based APR~\cite{liu2019tbar} performs well for known bugs but is powerless against previously unseen bugs. 
In contrast, learning-based APR~\cite{chen2022neural} is able to learn the bug-fixing patterns automatically from existing code repositories, thus yielding good generalization capability. 
Recently, as a further development of learning-based APR, large language model (LLM) based APR is receiving increasing attention from both academia and industry, delivering state-of-the-art performance~\cite{zhang2024systematic}.

ChatGPT~\cite{openai2023optimizing}, widely recognized as a leading application of LLMs, has been successfully integrated into various code-related tasks~\cite{sun2023automatic,zhang2023surveyllm}.
In the APR field, ChatGPT stands as the most widely-used LLM, and has facilitated a significant amount of subsequent APR work, greatly advancing the progress in APR~\cite{zhang2023gamma}.
Sobania et al. ~\cite{sobania2023analysis} evaluate the bug-fixing capabilities of ChatGPT and find ChatGPT is able to fix 31 out of 40 bugs on QuixBugs~\cite{lin2017quixbugs} benchmark. Xia et al.~\cite{xia2023keep} employ ChatGPT in a conversational manner to fix 114 and 48 bugs on the Defects4J-v1.2 and Defects4J-v2.0 benchmark.
Zhang et al.~\cite{zhang2023critical} demonstrate that ChatGPT is able to fix 109 out of 151 buggy programming problems.

However, during the continuous advancements of GPT-family models, few studies have systematically compared the effectiveness and differences across various GPT versions in APR. 
Typically, upgrades in GPT models are attributed to newer versions having more parameters and larger training datasets, thus resulting in better code understanding and improved APR performance.
Such a perspective may be challenged with the occurrence of the latest version of GPT, i.e., GPT-o1. 
Unlike its predecessors, GPT-o1 incorporates reinforcement learning (RL) and chain of thought (COT) techniques.
Particularly, before answering a question, it first takes time to think, organizes a chain of thought, and then arrives at a final answer.
By practice, O1 is more suitable for fields with complex logic and relatively definitive answers, such as mathematics, programming, and bug-fixing.

In this work, inspired by the fundamental differences between GPT-o1 and its predecessors, we conduct an empirical study to evaluate GPT-o1's performance in APR in a dialogue manner. 
We conduct experiments on the two currently available trial versions of the O1 model (i.e., O1-mini and O1-preview), as well as GPT-4o and ChatGPT, representing the most popular versions of the GPT-family models.
We also include other LLMs (e.g., CodeX) and APR techniques (e.g., CIRCLE) as baselines on the QuixBugs benchmark.
To this end, we design a two-step repair process. 
First, a basic prompt template is provided to the GPT models to directly repair the bugs.
For bugs that fail the test cases, we further provide models with the error information from the dynamic execution to perform a second round of repair. 

After all repairs are done, we collect data on the repair success rate, time spent on repairs, response length and behavior patterns of the model during the repair process, to further analyze O1's performance in APR. 
The results demonstrate that O1 outperforms ChatGPT (31/40) and GPT-4o (38/40) in APR, successfully repairing all 40 bugs in the Quixbugs benchmark. 
Moreover, O1's unique chain-of-thought reasoning pattern has proven effective for APR tasks.
By forming a chain-of-thought, O1 can better understand the logic of the buggy code, provide repair ideas, and deliver the correct repair code.
These findings are valuable and merit further attention in future research.

\section{Methodology}
In this section, we present the method for evaluating GPT-family models' ability to fix bugs. We use different GPT-family models to fix benchmark bugs and evaluate the repair behavior from multiple dimensions.

\subsection{Two-step Fix} \label{two-step-fix}
The process of bug fixing by GPT-family models consists of two steps. First, we use a basic repair template to ask the model whether there are bugs in the target program. After making a judgment, the model is requested to fix the bugs if they exist.

\begin{lstlisting}[language=pattern,caption={A basic template used to fix program bugs.},label={listing:bit-count-template}]
Does this program have a bug? How to fix 
it?

[Code of the Buggy Program]
\end{lstlisting}

Listing~\ref{listing:bit-count-template} shows a basic template we use to fix program bugs. We begin the prompt with the question \textit{Does this program have a bug? How to fix it?} to request GPT model to fix the following Python program, leaving a blank line between the question and the program.

After GPT model is asked to fix the buggy program, it will provide an initial fix answer. Previous GPT models (GPT-4o and earlier) would directly provide an answer, while the new version (ChatGPT-o1) will first spend some time thinking, develop a chain of thought, and then fix the program based on the chain of thought before presenting the complete fix code. 

We verify the correctness of the answer provided by GPT models with existing test set. For buggy programs that do not pass tests, we perform a secondary fix. To help GPT models better understand the reasons for the program errors, we include the error messages in the template used for the secondary fix. Listing~\ref{listing:sec-fix-template} is the prompt template used in the secondary fix step. Following \textit{The given corrected version fails to pass the test cases, and the results are as follows:}, we provide the original error report to prompt GPT models for further repair.

\begin{lstlisting}[language=pattern,caption={The secondary fix template for bug fixing.},label={listing:sec-fix-template}]
The given corrected version fails to 
pass the test cases, and the results are 
as follows:

[Error Message of Test Cases]

\end{lstlisting}

\subsection{Evaluation Metrics}
We evaluate the performance of GPT-o1 and prior GPT-family models in fixing program bugs from the following dimensions.

\begin{itemize}
    \item \textbf{Repair success rate}. We utilize available test cases to identify generated patch correctness, which is the most important metric for evaluating the effectiveness of APR techniques in the literature.
    
    \item \textbf{Time for thinking}. Compared to earlier versions of GPT models, O1 has added a "thinking" phase. 
    Thus, we record and analyze the time spent by O1-preview and O1-mini in thinking phase.
    
    \item \textbf{Response length}. We record and compare the output lengths of different models to evaluate the monetary cost of repairing.
    
    \item \textbf{Model behavior pattern}. We analyze behavior patterns of O1 and other GPT models to explore their mind when fixing bugs.
    
\end{itemize}

\subsection{Benchmark and Baselines}

We use QuixBugs~\cite{lin2017quixbugs} as the benchmark, which is widely adopted in the APR literature~\cite{zhang2023survey}.
For each of the 40 benchmark problems, we take the erroneous Python version. 
These programs have complete context and corresponding test cases, and they are relatively short, making them suitable for bug fixing through dialogue with GPT models.
We compare our results with previous work~\cite{sobania2023analysis,yuan2022circle}. 
In prompt designing, we keep consistent with previous work~\cite{sobania2023analysis} and carefully review the comments in the original program to ensure that they do not reveal the solution, retaining only the relevant parts of the test cases. 
Thus, we integrate the baseline method into the first step of the repair method presented in Section~\ref{two-step-fix}.

\section{Results}
In this section, we first compare the repair results of O1, previous GPT-family models, and the baseline method on QuixBugs. We then analyze and summarize the details of the O1's repair behavior.

\subsection{Comparison of O1, Previous GPT-family Models, and APR techniques}
\begin{table*}[ht!]
  \centering
  \caption{Results achieved by O1-preview, O1-mini, GPT-4o, ChatGPT, Codex~\cite{prenner2022can}, CIRCLE~\cite{yuan2022circle}, and the standard APR approaches~\cite{ye2021comprehensive} on the problems from the QuixBugs benchmark. For ChatGPT, the number of successful runs are listed in brackets. The baseline method~\cite{sobania2023analysis} also mentions that when additional information is given, the repair effectiveness of ChatGPT improves. We highlight these results in blue. For O1-preview, O1-mini and GPT-4o, we provide the failed test case output in the prompt and ask the model to try fixing it again. We mark the results of the second fix in red.}
\renewcommand{\arraystretch}{1.3}
  \label{tab:main-results}
  \resizebox{\textwidth}{!}{%
  \begin{tabular}{lccccccc}
    \hline
    \textbf{Benchmark problem}\;\;\;&
    \;\;\;\textbf{O1-preview}\;\;\;& 
    \;\;\;\textbf{O1-mini}\;\;\;& 
    \;\;\;\textbf{GPT-4o}\;\;\;
    &\textbf{ChatGPT}\;\;\; & \;\;\;
    \textbf{Codex~\cite{prenner2022can}}\;\;\; & \;\;\;
    \textbf{CIRCLE~\cite{yuan2022circle}}\;\;\; & \;\;\;
    \textbf{Standard APR~\cite{ye2021comprehensive}}\;\;\; \\
    \hline
    
    bitcount                 & \Checkmark & \Checkmark & \Checkmark & \XSolidBrush (0 / 4) \textcolor{blue}{\Checkmark} & \Checkmark & \Checkmark  & \XSolidBrush \\
    \rowcolor{lightgray}
    breadth-first-search    & \Checkmark  & \Checkmark & \Checkmark &  \Checkmark (2 / 4)  & \XSolidBrush & \Checkmark & \XSolidBrush \\
    bucketsort              & \Checkmark  & \Checkmark & \Checkmark &  \Checkmark (4 / 4) & \Checkmark & \Checkmark & \XSolidBrush \\
    \rowcolor{lightgray}
    depth-first-search      & \Checkmark  & \Checkmark & \Checkmark &  \XSolidBrush (0 / 4) \textcolor{blue}{\Checkmark} & \Checkmark & \XSolidBrush & \XSolidBrush \\
    detect-cycle            & \Checkmark  & \Checkmark & \Checkmark &  \XSolidBrush (0 / 4) \textcolor{blue}{\Checkmark} & \XSolidBrush & \Checkmark & \Checkmark \\
    \rowcolor{lightgray}
    find-first-in-sorted    & \Checkmark  & \Checkmark & \Checkmark &  \Checkmark (2 / 4) & \Checkmark & \Checkmark & \XSolidBrush \\
    find-in-sorted          & \Checkmark  & \Checkmark & \Checkmark &  \Checkmark (3 / 4) & \XSolidBrush & \Checkmark & \XSolidBrush \\
    \rowcolor{lightgray}
    flatten                 &  \Checkmark & \Checkmark & \Checkmark &  \Checkmark (4 / 4) & \Checkmark & \Checkmark & \XSolidBrush \\
    gcd                     & \Checkmark  & \Checkmark & \XSolidBrush \textcolor{red}{\Checkmark} &  \XSolidBrush (0 / 4) \textcolor{blue}{\Checkmark}& \Checkmark & \Checkmark & \XSolidBrush \\
    \rowcolor{lightgray}
    get-factors             & \Checkmark  & \Checkmark & \Checkmark &  \Checkmark (1 / 4) & \Checkmark & \Checkmark & \XSolidBrush \\
    hanoi                   &  \Checkmark & \Checkmark & \XSolidBrush \textcolor{red}{\XSolidBrush} &  \XSolidBrush (0 / 4) \textcolor{blue}{\Checkmark}& \Checkmark & \Checkmark & \XSolidBrush \\
    \rowcolor{lightgray}
    is-valid-parenthesization& \Checkmark & \Checkmark & \Checkmark &  \Checkmark (2 / 4) & \Checkmark & \XSolidBrush & \XSolidBrush \\
    kheapsort               & \Checkmark  & \Checkmark & \Checkmark &  \XSolidBrush (0 / 4) \textcolor{blue}{\XSolidBrush}& \Checkmark & \XSolidBrush & \XSolidBrush \\
    \rowcolor{lightgray}
    knapsack               & \Checkmark   & \Checkmark & \Checkmark &  \Checkmark (1 / 4) & \Checkmark & \Checkmark & \Checkmark \\
    kth                     & \Checkmark  & \Checkmark & \Checkmark &  \XSolidBrush (0 / 4) \textcolor{blue}{\Checkmark}& \XSolidBrush & \XSolidBrush & \XSolidBrush \\
    \rowcolor{lightgray}
    lcs-length             &  \Checkmark  & \Checkmark & \Checkmark &  \XSolidBrush (0 / 4) \textcolor{blue}{\XSolidBrush}& \XSolidBrush & \Checkmark & \XSolidBrush \\
    levenshtein            & \Checkmark   & \Checkmark & \Checkmark &  \XSolidBrush (0 / 4) \textcolor{blue}{\Checkmark} & \XSolidBrush & \Checkmark & \Checkmark \\
    \rowcolor{lightgray}
    lis                   &  \XSolidBrush \textcolor{red}{\Checkmark}   & \XSolidBrush \textcolor{red}{\Checkmark} & \Checkmark &  \XSolidBrush (0 / 4) \textcolor{blue}{\XSolidBrush}& \XSolidBrush & \XSolidBrush & \Checkmark \\
    longest-common-subsequence & \Checkmark & \Checkmark & \Checkmark &  \XSolidBrush (0 / 4) \textcolor{blue}{\XSolidBrush}& \Checkmark & \XSolidBrush & \XSolidBrush \\
    \rowcolor{lightgray}
    max-sublist-sum        &  \Checkmark  & \XSolidBrush \textcolor{red}{\Checkmark} & \XSolidBrush \textcolor{red}{\Checkmark} &  \XSolidBrush (0 / 4) \textcolor{blue}{\Checkmark} & \Checkmark & \XSolidBrush & \XSolidBrush \\
    mergesort             & \Checkmark    & \Checkmark & \Checkmark &  \Checkmark (1 / 4) & \XSolidBrush & \Checkmark & \Checkmark \\
    \rowcolor{lightgray}
    minimum-spanning-tree  & \Checkmark   & \Checkmark & \Checkmark &  \XSolidBrush (0 / 4) \textcolor{blue}{\Checkmark}& \XSolidBrush & \Checkmark & \XSolidBrush \\
    next-palindrome       &  \Checkmark   & \Checkmark & \Checkmark &  \Checkmark (1 / 4) & \XSolidBrush & \Checkmark & \XSolidBrush \\
    \rowcolor{lightgray}
    next-permutation    &  \Checkmark     &\Checkmark & \Checkmark &  \XSolidBrush (0 / 4) \textcolor{blue}{\Checkmark}& \XSolidBrush & \Checkmark & \XSolidBrush \\
    pascal             & \Checkmark       & \Checkmark & \Checkmark & \Checkmark (1 / 4) & \XSolidBrush & \Checkmark & \XSolidBrush \\
    \rowcolor{lightgray}
    possible-change      &   \Checkmark   & \Checkmark & \Checkmark &  \Checkmark (1 / 4) & \Checkmark & \XSolidBrush & \XSolidBrush \\
    powerset             &    \Checkmark  & \Checkmark & \Checkmark &  \XSolidBrush (0 / 4) \textcolor{blue}{\Checkmark}& \Checkmark & \Checkmark & \XSolidBrush \\
    \rowcolor{lightgray}
    quicksort           &   \Checkmark    & \Checkmark & \Checkmark &  \Checkmark (1 / 4) & \Checkmark & \Checkmark & \Checkmark \\
    reverse-linked-list   &  \Checkmark   & \Checkmark & \Checkmark &  \Checkmark (2 / 4) & \Checkmark & \XSolidBrush & \XSolidBrush \\
    \rowcolor{lightgray}
    rpn-eval              &   \Checkmark  & \Checkmark & \Checkmark &  \XSolidBrush (0 / 4) \textcolor{blue}{\XSolidBrush}& \XSolidBrush & \XSolidBrush & \Checkmark \\
    shortest-path-length   & \XSolidBrush \textcolor{red}{\Checkmark}   & \XSolidBrush \textcolor{red}{\Checkmark} & \XSolidBrush \textcolor{red}{\Checkmark} &  \Checkmark (1 / 4) & \XSolidBrush & \XSolidBrush & \XSolidBrush \\
    \rowcolor{lightgray}
    shortest-path-lengths  &  \Checkmark  & \Checkmark & \Checkmark & \XSolidBrush (0 / 4) \textcolor{blue}{\XSolidBrush}& \XSolidBrush & \XSolidBrush & \XSolidBrush \\
    shortest-paths       &   \Checkmark   & \Checkmark & \Checkmark &  \Checkmark (1 / 4) & \XSolidBrush & \XSolidBrush & \XSolidBrush \\
    \rowcolor{lightgray}
    shunting-yard        &   \Checkmark   & \Checkmark & \Checkmark &  \Checkmark (2 / 4) & \XSolidBrush & \XSolidBrush & \XSolidBrush \\
    sieve                &  \Checkmark    & \Checkmark & \Checkmark &  \XSolidBrush (0 / 4) \textcolor{blue}{\Checkmark}& \Checkmark & \Checkmark & \XSolidBrush \\
    \rowcolor{lightgray}
    sqrt                 &  \Checkmark    & \Checkmark & \Checkmark &  \Checkmark (1 / 4) & \Checkmark & \XSolidBrush & \XSolidBrush \\
    subsequences        &  \Checkmark     & \Checkmark & \Checkmark &  \Checkmark (1 / 4) & \XSolidBrush & \Checkmark & \XSolidBrush \\
    \rowcolor{lightgray}
    to-base            &    \Checkmark    & \Checkmark & \Checkmark &  \XSolidBrush (0 / 4) \textcolor{blue}{\XSolidBrush} & \Checkmark & \XSolidBrush & \XSolidBrush \\
    topological-ordering  & \Checkmark    & \Checkmark & \Checkmark &  \XSolidBrush (0 / 4) \textcolor{blue}{\XSolidBrush}& \XSolidBrush & \Checkmark & \XSolidBrush \\
    \rowcolor{lightgray}
    wrap                &   \Checkmark    & \Checkmark & \XSolidBrush \textcolor{red}{\XSolidBrush} &  \XSolidBrush (0 / 4) \textcolor{blue}{\XSolidBrush}& \Checkmark & \XSolidBrush & \XSolidBrush \\
    \hline
    \pmb{$\Sigma$}\textbf{ (Solved)} & \textbf{38(\textcolor{red}{40})} & \textbf{37(\textcolor{red}{40})} & \textbf{35(\textcolor{red}{38})} &  \textbf{19(\textcolor{blue}{31})}   & \textbf{21}  & \textbf{23} & \textbf{7} \\
  \hline 
\end{tabular}
}
\end{table*}
We present the main results of comparison in Table~\ref{tab:main-results}. In terms of baseline results, despite the significant improvement in repair effectiveness after introducing additional information, with a repair success rate reaching 31/40, it is still far from latest GPT models (GPT-4o, O1-mini, O1-preview). It can be seen that GPT-family models have made significant improvements in program repair capabilities through iterations.

Compared to the current mainstream GPT model (GPT-4o), O1 shows some improvement in program repair capabilities. Before providing test case error information, O1-mini and O1-preview can repair 37 and 38 bugs respectively, which is 2 and 3 more than GPT-4o. After providing test case error information, both O1-mini and O1-preview are able to repair all 40 bugs, whereas GPT-4o can only repair 38 bugs.

To further investigate the improvements of the O1 model compared to other GPT-family models in program repair, we also conduct a case analysis on the programs that the O1 model is able to repair but GPT-4o and previous GPT models can not. We find that these programs are relatively complex, involving recursion and nested loops, and more relative to real-world problems. Take hanoi as instance, both O1-preview and O1-mini successfully repair the bug on the first attempt, whereas neither GPT-4o nor ChatGPT are able to fix it. ChatGPT fails four times util the correct test case answers are provided. GPT-4o incorrectly interpret the boundary condition, assuming that the source rod cannot be equal to the destination rod. In order to solve the hanoi problem, the O1 model spends 15 seconds thinking and forms a chain of thought (analyze functionality, check code logic, correct steps, optimize move steps). This chain of thought helps it correctly understand the logic of the problem, avoiding falling into incorrect logical branches. From this, we can infer that the reasoning pattern from chain of thought to solution is crucial for repairing buggy programs with complex logic.

\subsection{Evaluation of O1 on Response Time, Response Length, and Behavior Pattern}
In terms of response length, we conduct a token-count analysis of O1-preview, O1-mini, and GPT-4o on all benchmark programs with tiktoken\footnote{https://github.com/openai/tiktoken}. Full results can be found in Table~\ref{tab:token-length}. We find that the average response lengths for the three models are 1450, 1086, and 654 tokens, respectively. This indicates that the O1's response length is over 50\% longer compared to previous GPT models, leading to more comprehensive and complete answers, albeit at a higher cost. 

In terms of response time, since the generation time is related to the length of output, we only measure the time taken by O1 for thinking. Full results can be found in Table~\ref{tab:think-time}. We find that the average thinking time for O1-preview reaches 19.82 seconds, which is about three times longer than that of O1-mini (7.02 s). Due to GPT-4o's ability to generate outputs directly without a "thinking" phase, Table~\ref{tab:think-time} does not include GPT-4o when calculating thinking time. Considering the relationship between generation time and the output length of the model, we estimate that O1 takes over 70\% more time on program repair tasks compared to previous GPT models. Thus when repairing relatively simple programs without involving too much context and complex logical relationships, considering the costs of time and money, traditional GPT models seem to be more practical compared to O1.

\begin{figure}[htp]
  \centering
  \includegraphics[width=0.5\textwidth]{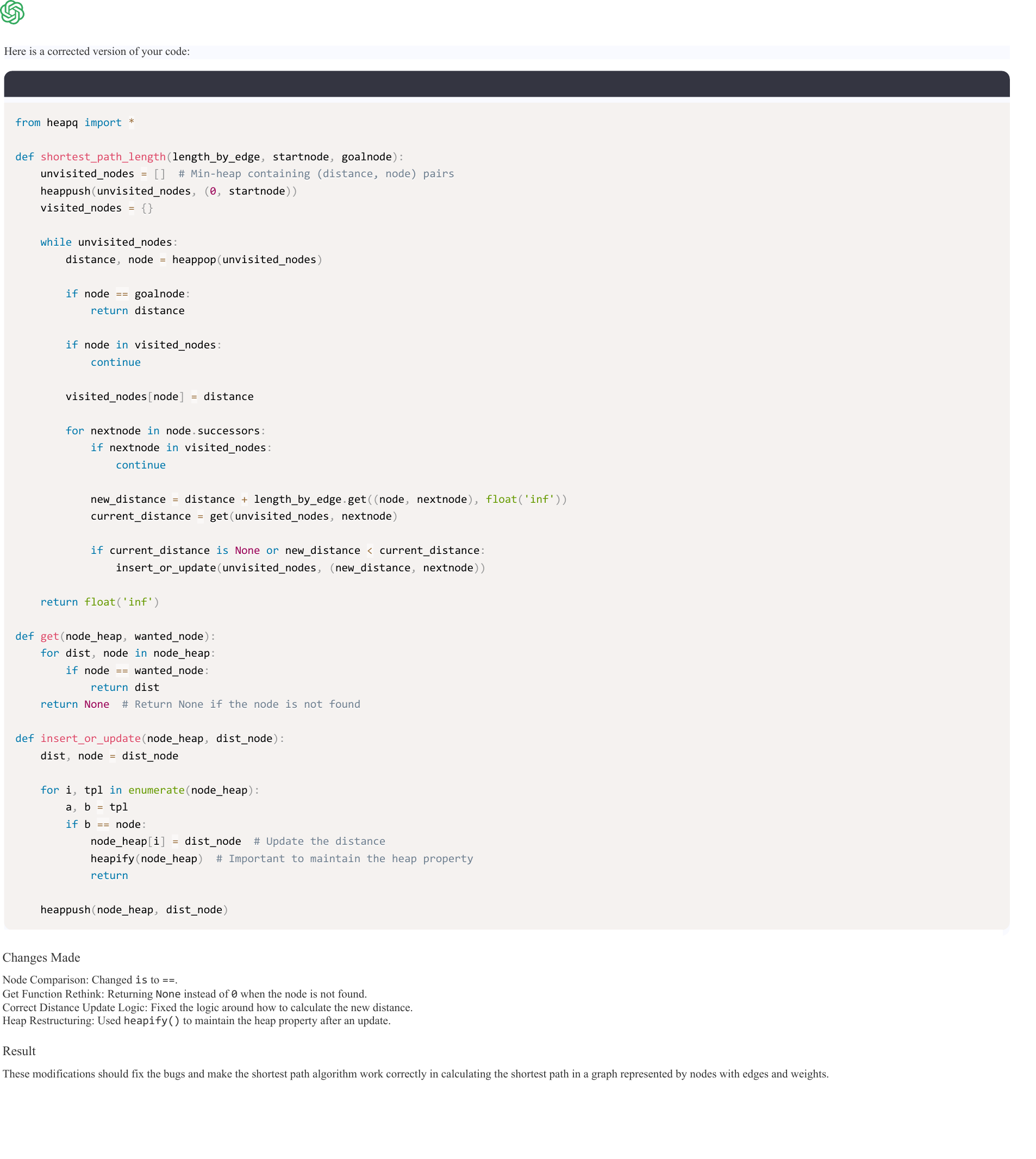}
  \caption{Behavior patterns of GPT-4o, using the chat history when solving the problem shortest-path-length as an example.}
  \label{fig:chat-history-4o}
\end{figure}

\begin{figure*}[htp]
  \centering
  \includegraphics[width=\textwidth]{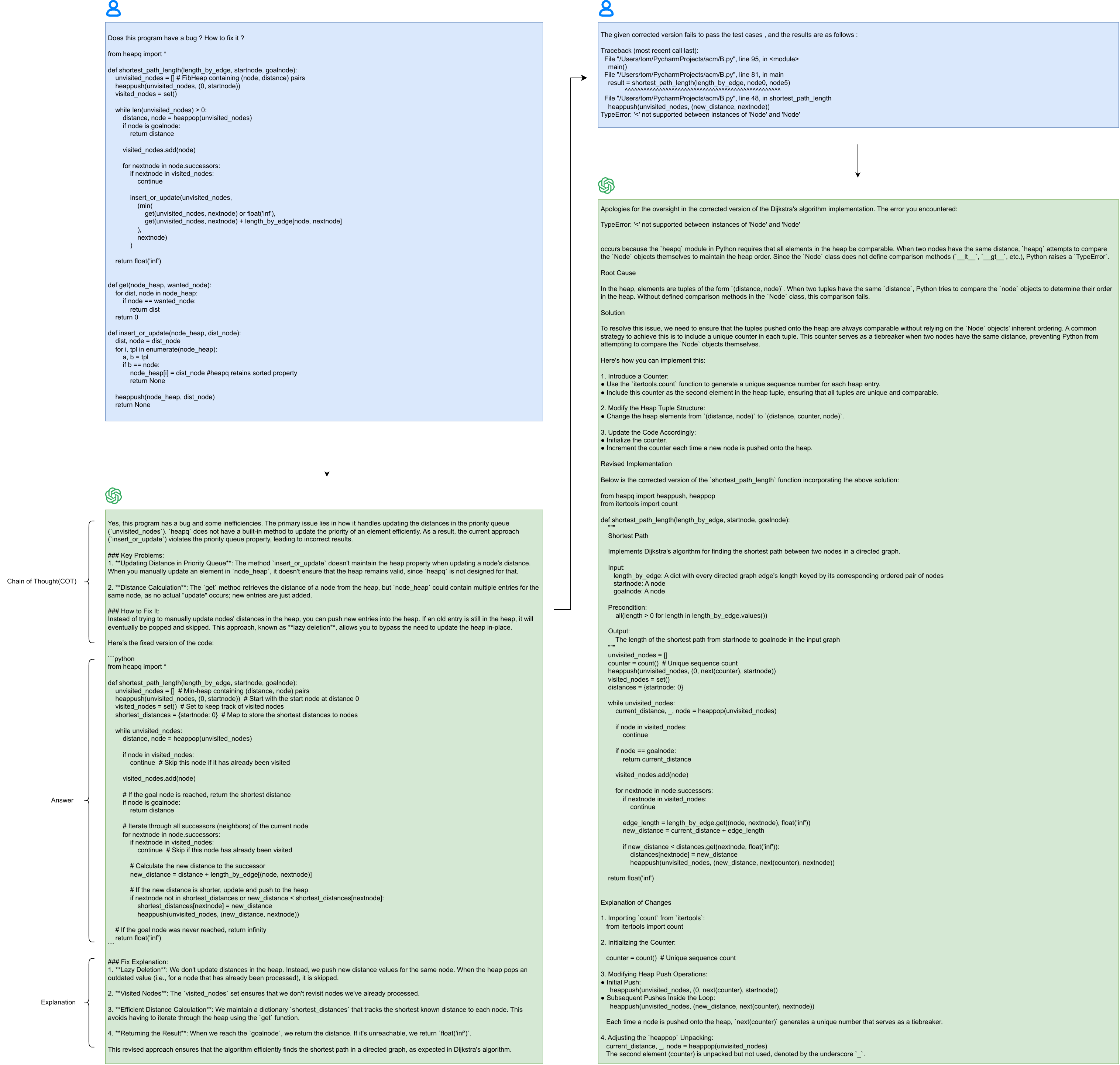}
  \caption{Behavior patterns of O1, using the chat history when solving the problem shortest-path-length as an example.}
  \label{fig:chat-history}
\end{figure*}

In terms of model behavior patterns, we observe that O1 typically begins by performing logical analysis on the buggy program, generating a solution, and then gradually proceeding with repairs and testing before providing the complete fixed code. In contrast, GPT-4o tends to provide the repaired code first, followed by an explanation of the code, as shown in Figure \ref{fig:chat-history-4o}. In order to illustrate the behavior patterns and response styles of O1 more specifically, we conduct a detailed analysis of O1's thought process when addressing the shortest-path-length problem. As shown in Figure \ref{fig:chat-history}, we first use the repair template to ask O1 how to fix the defects in the shortest-path-length problem. After careful consideration, O1 provides a response that consists of three parts (Chain of Thought (COT), Answer, Explanation). The part of chain of thought (COT) identifies that the key problems in the flawed program lie in two areas which corresponding to updating the distance in the priority queue and issues with distance calculation. Then, O1 provides corrective suggestions based on the identified problems. In the second part, O1 provides the complete revised version of the program. In the third part, O1 provides an explanation of the code presented in the second part, explaining it in the order of the code statements. O1 sequentially describes the functionality of each section of the code, which is similar to code comments. The same behavior pattern can also be observed in the repair during the second step. We provide the error message to O1 so that it can correct the previous wrong answer. In the process of self correction, O1 summarizes the underlying reasons for its incorrect answers and provides solutions. It then makes a second revision based on its previous answer and presents what it believes to be the final version. Following this, it also explains the changes made to the code. Based on such observation, we believe that the behavior pattern of O1 is overall consistent and aligns with human thought processes. When solving a problem, it first analyzes the reasons and provides a general idea, then offers a specific solution based on that idea. Afterward, it reflects on and explains its own response.


\begin{table}[ht]
    \centering
    \caption{The token length output by O1 and GPT-4o when fixing each benchmark problem.}
    \resizebox{0.5\textwidth}{!}{%
    \begin{tabular}{lccc}
        \toprule
        \textbf{Benchmark Problem} & \textbf{O1-preview} & \textbf{O1-mini} & \textbf{GPT-4o} \\
        \midrule
        bitcount                   & 1846 & 1172    & 647  \\
        breadth-first-search       & 1179 & 1015    & 780  \\
        bucketsort                 & 1540 & 710     & 669  \\
        depth-first-search         & 1620 & 738     & 667  \\
        detect-cycle               & 1518 & 719     & 757  \\
        find-first-in-sorted       & 2099 & 1406    & 820  \\
        find-in-sorted             & 2159 & 912     & 785  \\
        flatten                    & 987  & 856     & 487  \\
        gcd                        & 1393 & 822     & 456  \\
        get-factors                & 1109 & 932     & 553  \\
        hanoi                      & 1826 & 1262    & 726  \\
        is-valid-parenthesization  & 878  & 578     & 770  \\
        kheapsort                  & 1282 & 1203    & 1078 \\
        knapsack                   & 1722 & 1029    & 904  \\
        kth                        & 1848 & 1171    & 472  \\
        lcs-length                 & 1345 & 927     & 589  \\
        levenshtein                & 1485 & 914     & 727  \\
        lis                        & 1612 & 896     & 597  \\
        longest-common-subsequence & 1299 & 875     & 710  \\
        max-sublist-sum            & 1423 & 1370    & 485  \\
        mergesort                  & 575  & 1017    & 1438 \\
        minimum-spanning-tree      & 1386 & 1346    & 903  \\
        next-palindrome            & 1215 & 1057    & 1658 \\
        next-permutation           & 1642 & 931     & 817  \\
        pascal                     & 1778 & 798     & 692  \\
        possible-change            & 940  & 860     & 798  \\
        powerset                   & 1642 & 1128    & 563  \\
        quicksort                  & 760  & 691     & 281  \\
        reverse-linked-list        & 1062 & 1005    & 342  \\
        rpn-eval                   & 1638 & 1097    & 415  \\
        shortest-path-length       & 814  & 1097    & 768  \\
        shortest-path-lengths      & 2035 & 1263    & 432  \\
        shortest-paths             & 1130 & 1244    & 495  \\
        shunting-yard              & 2104 & 1329    & 573  \\
        sieve                      & 1432 & 1067    & 511  \\
        sqrt                       & 931  & 812     & 268  \\
        subsequences               & 1557 & 653     & 363  \\
        to-base                    & 968  & 1015    & 281  \\
        topological-ordering       & 1497 & 2755    & 390  \\
        wrap                       & 1374 & 752     & 383  \\
        \midrule
        \textbf{Average}             & 1450.07 & 1086.90 & 654.07  \\
        \bottomrule
    \end{tabular}}
    \label{tab:token-length}
\end{table}
\begin{table}[ht]
    \centering
    \caption{The thinking time spent on repairing each benchmark problem using the O1 model. When thinking time is less than five seconds, we count it as five seconds.}
    \resizebox{0.5\textwidth}{!}{%
    \begin{tabular}{lccc}
        \toprule
        \textbf{Benchmark problem} & \textbf{O1-preview} & \textbf{O1-mini} & \textbf{GPT-4o} \\
        \midrule
        bitcount & 17s & < 5s & \textasciitilde \\
        breadth-first-search & 27s & < 5s & \textasciitilde \\
        bucketsort & 18s & < 5s & \textasciitilde \\
        depth-first-search & 9s & < 5s & \textasciitilde \\
        detect-cycle & 23s & < 5s & \textasciitilde \\
        find-first-in-sorted & 31s & 11s & \textasciitilde \\
        find-in-sorted & 18s & < 5s & \textasciitilde \\
        flatten & 19s & < 5s & \textasciitilde \\
        gcd & 8s & < 5s & \textasciitilde \\
        get-factors & 16s & < 5s & \textasciitilde \\
        hanoi & 15s & 7s & \textasciitilde \\
        is-valid-parenthesization & 10s & < 5s & \textasciitilde \\
        kheapsort & 30s & 6s & \textasciitilde \\
        knapsack & 38s & 5s & \textasciitilde \\
        kth & 28s & 10s & \textasciitilde \\
        lcs-length & 12s & < 5s & \textasciitilde \\
        levenshtein & 18s & < 5s & \textasciitilde \\
        lis & 51s & 38s & \textasciitilde \\
        longest-common-subsequence & 15s & 42s & \textasciitilde \\
        max-sublist-sum & 21s & 6s & \textasciitilde \\
        mergesort & 17s & < 5s & \textasciitilde \\
        minimum-spanning-tree & 28s & 10s & \textasciitilde \\
        next-palindrome & 53s & 8s & \textasciitilde \\
        next-permutation & 23s & 7s & \textasciitilde \\
        pascal & 33s & 8s & \textasciitilde \\
        possible-change & 14s & < 5s & \textasciitilde \\
        powerset & 23s & < 5s & \textasciitilde \\
        quicksort & 11s & < 5s & \textasciitilde \\
        reverse-linked-list & 14s & < 5s & \textasciitilde \\
        rpn-eval & 16s & 5s & \textasciitilde \\
        shortest-path-length & 16s & 7s & \textasciitilde \\
        shortest-path-lengths & 13s & < 5s & \textasciitilde \\
        shortest-paths & 21s & < 5s & \textasciitilde \\
        shunting-yard & 17s & < 5s & \textasciitilde \\
        sieve & 15s & 8s & \textasciitilde \\
        sqrt & 10s & 5s & \textasciitilde \\
        subsequences & 47s & < 5s & \textasciitilde \\
        to-base & 20s & < 5s & \textasciitilde \\
        topological-ordering & 18s & 6s & \textasciitilde \\
        wrap & 21s & < 5s & \textasciitilde \\
        \midrule
        \textbf{Average} & 19.82s & 7.02s & \textasciitilde \\
        \bottomrule
    \end{tabular}}
    \label{tab:think-time}
\end{table}

\section{Related Works}
Traditional APR methods can be mainly categorized into search-based~\cite{Simfix:2018,10.1145/3468264.3468600,8823627}, constraint-based~\cite{196,106,114}, and template-based~\cite{44,45,167short} approaches. 
Search-based methods are also known as heuristics solutions. The core idea of search-based
methods is to search for the correct patch in a predefined patch space. Search-based
methods first look in the search space to identify the most likely locations of buggy code in the program, usually with the help of heuristics to generate candidate fixes, and then search for suitable fix patches by means of mutation-selection, test execution, traversal strategies, and etc. As an early class of APR techniques, it has provided preliminaries for the generation of other repair methods. On the other hand, search-based methods have also exposed some problems, such as the explosive nature of the search space and the patch overfitting problem \cite{10.1145/3318162,DBLP:journals/corr/abs-1807-00515}. These problems have been continuously improved upon by later methods.

Constraint-based methods guides the repair process
by developing a set of constraint specifications. Such techniques transform the program repair problem into a constraint solver problem and use formal specifications to quickly prune infeasible parts to find expression-level variations that satisfy constraints collected from tests by program analysis, thus facilitating patch generation. Among constraint-based methods, a lot of innovations have been made in semantic works (syntactic-semantic combination \cite{9438564,10.1145/3106237.3106309}, formal specifications\cite{10.1145/3418461,10.1145/3180155.3180250}), all of which are revolutionary in the APR process. Compared with search-based heuristics, constraint-based methods do not require extensive search and therefore have much lower computational resources and costs and higher repair efficiency. However, such methods rely heavily on the formulated specification constraints to guide the program repair, so it lacks flexibility. 

Template-based methods utilize a predefined program fix template \cite{10.5555/2486788.2486893,10.1145/3180155.3180245,44} to generate repair patches. The fix templates may be manually extracted (e.g., kPAR \cite{DBLP:journals/corr/abs-1812-07283}) or automatically mined (e.g., HDRepair \cite{7476644}). After fault localization, the template-based APR tool will target these defects to select the corresponding fix templates to generate candidate patches. Recent study \cite{Liu_2020} shows that template-based repair methods are more effective compared to search-based and constraint-based methods. However, template-based methods cannot repair defects that are not summarized by templates, which means that defects that have not been encountered exceed the repair capabilities of template-based methods.

In summary, traditional repair methods, although effective, struggle to achieve a balance between performance and effectiveness. Compared with traditional APR methods, learning-based~\cite{VulRepair,49,TransR,zhang2023pre} APR leverages prior knowledge from pre-trained models to automatically fix program defects, demonstrating wide-ranging effectiveness. It is capable of cross-language repairs and can correct previously unseen defects. Building on this foundation, researchers~\cite{10172803,xia2023conversationalautomatedprogramrepair,zhang2024no} further utilize LLMs to further enhance the repair effects. Fine-tuning methods~\cite{Berabi2021TFixLT,9699412,chi2022seqtransautomaticvulnerabilityfix} finetune the pre-trained LLMs on the bug, vulnerability, or error repair dataset so that the LLMs can obtain the corresponding repair knowledge and work effectively on the downstream APR task. Compared with the fine-tuning methods, prompt learning methods~\cite{fan2023automatedrepairprogramslarge,joshi2022repairnearlygenerationmultilingual,pearce2022examiningzeroshotvulnerabilityrepair} emphasize the LLM’s code understanding and generation capabilities. Both methods demonstrate wide application and great effectiveness.

To further study the capability of LLMs on APR task, this paper conducts extensive research on one of the most representative LLMs (GPT-family models) for APR. Based on previous work~\cite{sobania2023analysis,zhang2023critical} and in conjunction with the newly released GPT model O1, we provide a comprehensive analysis and comparison of the performance and differences of representative models within the GPT-family in the context of APR.

\section{Conclusion}
This paper conducts the first empirical study to evaluate the capabilities of the latest GPT-o1 model in automated program repair. 
The experimental results on the QuixBugs benchmark demonstrate the advancements of GPT-o1 against previous GPT-family models and repair techniques.
We also provide a comprehensive analysis of GPT-o1 from four dimensions, like thinking time and response length.
Besides, we conduct a case analysis and conclude the advantages of GPT-o1 in repairing complex bugs due to chain of thought. 
Our findings offer a reference for future in-depth research on utilizing GPT-o1 and GPT models for program repair.

\section{Limitations}
Currently, O1 is in the trial phase, with usage limits and high API costs. Understanding of O1 is still quite insufficient. The benchmark used in our study is relatively small, and further research is needed with larger datasets to explore the O1 model's capabilities in APR more comprehensively. 

\bibliographystyle{IEEEtran}
\bibliography{custom}

\end{document}